\documentclass[aps,prb,twocolumn,showpacs]{revtex4-1}%
\usepackage{amsfonts}
\usepackage{amsmath}
\usepackage{amssymb}
\usepackage[pdftex]{graphicx}
\usepackage{dcolumn}
\usepackage{bm}
\begin{document}
\date{\today}
%LA-UR 94-
\title
{London Penetration Depth and Pair Breaking}
\author{V. G. Kogan, R. Prozorov }
\address{Ames Laboratory and Physics Department ISU, Ames, IA
50011}
\author{V. Mishra }
\address{Materials Science Division, Argonne National Laboratory, Lemont, IL-60439}

\begin{abstract}
The London penetration depth is evaluated for isotropic materials for any transport and pair-breaking Born scattering rates. Besides known results, a number of new features are found. The slope $|d\rho/d\theta |$ of the normalized superfluid density $\rho=\lambda^2(0)/\lambda^2(\theta)$ at the transition $\theta=T/T_c=1$  has a minimum near the value of the pair-breaking parameter  separating gapped and gapless states.  The low-$T$ exponentially flat part of $\rho$ for the s-wave materials is suppressed by increasing pair breaking. For   strong $T_c $ suppression by magnetic impurities the ``Homes scaling" $\lambda^{-2}(0)   \propto   \sigma T_c$ with $\sigma$ being the normal conductivity  gives way to $\lambda^{-2}(0)   \propto   \sigma T_c^2$.  
For the d-wave order parameter, the transport and spin-flip Born scattering rates enter the theory only as a sum, in particular, they affect the $T_c$ depression in the same manner. 
We confirm that the linear low temperature behavior  of $\rho$ in a  broad   range of the combined scattering parameter  turns to the $T^2$ behavior only when the critical temperature is suppressed at least by a factor of 3 relative to the clean limit $T_{c0}$. Moreover, in this range, $\rho (\theta)$ is only weakly dependent on the scattering parameter,  i.e. it is nearly universal.  
\end{abstract}
\maketitle

%%%%%%
\section{Introduction}
%%%%%%%%

Within isotropic weak coupling BCS theory, the London penetration depth $\lambda$ has been  evaluated for any concentration of non-magnetic impurities.\cite{AGD,nonloc,PK-ROPP} Spin-flip scattering suppresses the critical temperature $T_c$, the order parameter $\Delta$, and the superfluid density making   the analytic  evaluation of $\lambda$ more complicated. Early calculations of Abrikosov and Gor'kov (AG) for the strong pair breaking in the gapless state,\cite{AG}   of Skalski {\it et al} and  of Maki for the dirty limit \cite{Skalski,Maki-Parks} were done treating the scattering as weak and employing the Born approximation. 

Later developments related to cuprate d-wave superconductivity brought about a refined treatment of scattering, the t-matrix approach, the unitary limit.\cite{Muz88,Muz89,Muz94,Hirsch-Gold,Sun-Maki,Sauls} In particular, Hirshfeld and Goldenfeld showed how the unitary scattering may reconcile the $T^2$ low temperature dependence of $\lambda(T)-\lambda(0)$ with nearly impurity independent $T_c$ of some cuprates.\cite{Hirsch-Gold} 

Unprecedented number of new superconductors have been discovered since then,   the iron-based family of materials is rich in particular. Most of these materials have the pair-breaking scattering present to various degrees which are not yet clearly established.\cite{Kogan1,Kogan2,Kogan3} There are also examples of long mean-free-path paramagnets or antiferromagnets  which become superconducting with the temperature dependence of the penetration depth  corresponding to a strong pair breaking (CeCoIn$_5$). \cite{Kogan-Martin-Prozorov}
Hence, the necessity to have at least a qualitative picture of the pair breaking influence on $\lambda(T)$ in a broad range of scattering parameters.

Accurate methods for  measurements  of $\lambda(T)$, such as the tunnel-diod-resonator, are now available and used to extract information on the order parameter symmetry.  Å 
In this text, besides demonstrating substantial simplifications brought about by employing the quasi-classical   formalism\cite{E} to the problem of the penetration depth in materials with pair breaking, we provide a straightforward numerical procedure which can be used by those involved in studies of new materials and confronted with a need to estimate the contribution of pair breaking to various properties. 

For completeness of the presentation  we reproduce a number of well-known results for the s- and d-wave order parameters,  although the properties of the penetration depth are   our main goal and we try to present them in a form useful  to the community dealing with actual measurements. In particular, we focus on evaluation of the normalized superfluid density $\rho(T)=\lambda^2(0)/\lambda^2(T)$ and show that for the s-wave case, the slope of this quantity at $T\to T_c$ is a non-monotonic function of the pair breaking scattering parameter. We also show that the pair breaking suppresses the exponentially flat low temperature part of $\rho(T)$ at large scattering rates which can be confused with the d-wave behavior. 

For the d-wave order parameter, we find that in the Born approximation the linear low temperature behavior of $\lambda(T)$ is quite robust with respect to magnetic scattering and turns to  $\lambda\propto T^2$ only for strong pair breaking in agreement with conclusions of Ref.\,\onlinecite{Sauls}. This is in a striking difference with predictions based on unitary  scattering limit which suggests  that a small concentration of strong scatterers   destroys the d-wave low-$T$ linear signature of $\lambda(T)$.\cite{Hirsch-Gold,Sun-Maki}  We  find that in the Born approximation,  slopes of $\rho(t)=\lambda^2(0)/\lambda^2(\theta)$ at $T_c$ plotted vs reduced $\theta=T/T_c$ deviate  from the clean limit value of 4/3  only in materials with strongly suppressed $T_c$'s. Also, we 
find that for the isotropic Fermi surface the penetration depth for the d-wave order parameter $\Delta\propto  k_x^2-k_y^2$ is isotropic too ($\lambda_c=\lambda_{ab}$), although this is not required by symmetry. In fact, the anisotropy of $\lambda$  has been demonstrated for 
$\Delta\propto (k_x+ik_y)^2$ in Ref.\,\onlinecite{Muz89}. 

We also find that within our weak coupling model, the ratio $\Delta(0)/T_c$ (often taken as indicator for  the weak or strong coupling superconductivity), depends on the pair-breaking scattering and may exceed substantially the weak-coupling value  1.76.

%%%%%%
\section{Magnetic impurities }
%%%%%%%%

The system of equations describing superconductivity in this situation is:\cite{E}
\begin{eqnarray}
&&{\bm v} {\bm  \Pi}f=2  \Delta g   -2\omega f +\frac{g}{\tau_-}\langle f\rangle -
\frac{f}{\tau_+}\langle g\rangle \,,\label{Eil1}\\ 
&& -{\bm v} {\bm  \Pi}^*f^+=2  \Delta^* g   -2\omega f^+ + \frac{g}{\tau_-}\langle f^+\rangle -
\frac{f^+}{\tau_+}\langle g\rangle \,,\label{Eil2}\\ 
&& g^2=1-ff^{+}\,, \label{eil3}\\
 &&    \frac{\Delta}{2\pi T}\ln \frac{T_{c0}}{T}=\sum_{\omega >0}
\left (\frac{\Delta}{\hbar\omega}-\langle f\rangle\right )\,.
\label{eil4}\\
&&{\bm j}=-4\pi |e|N(0)T\,\, {\rm Im}\sum_{\omega >0}\langle {\bm v}g\rangle\,.
\label{eil5}
\end{eqnarray}
Here ${\bm v}$ is the Fermi velocity, ${\bm \Pi} =\nabla +2\pi i{\bm 
A}/\phi_0$; $\Delta({\bm r})$ is 
 the order parameter, $f({\bm r},{\bm v},\omega),\,\, 
f^{+}=f^*({\bm r},-{\bm v},\omega)$, and $g$ are Eilenberger Green's 
functions, $N(0)$ is the density of states at the Fermi level per spin; 
$\hbar\omega=\pi T(2n+1)$ with an integer $n$; $\langle ...\rangle$ 
stands for  averages over the Fermi surface;  $\bm j$ is the current density.

  The  scattering in the Born approximation is characterized  by two
scattering times, $\tau $ for the transport scattering responsible for
conductivity in the normal state, and   $\tau_m$ for the pair-breaking
magnetic scattering processes: 
  \begin{equation}
\frac{1}{\tau_\pm}=\frac{1}{\tau }\pm\frac{1}{\tau_m} \,.
  \label{rhos}
  \end{equation}

The self-consistency equation (\ref{eil4}) contains $T_{c0}$, the critical temperature in the absence of magnetic impurities.  This equation   is not always convenient because the actual $T_c\ne T_{c0}$ does not enter explicitly this form. It 
 can be recast to a form containing $T_c$, but our numerical procedure based on Eq.\,(\ref{eil4}) generates--among other things--the actual $T_c$ for a given $\tau_m$.
 
 %%%%%%%
\subsection{London penetration depth, s-wave}
%%%%%%%

We aim at finding how  weak fields penetrate the material, the problem  solved by perturbations. Hence, we have to solve first for the uniform zero-field state, we denote corresponding  functions as $f_0,g_0,\Delta_0$. In the isotropic situation of interest here, the sign of averages over the Fermi sphere can be omitted.  Equations (\ref{Eil1})-(\ref{eil4}) reduce to:
      \begin{eqnarray}
&&0=\sqrt{1-f_0^2}\,(\Delta_0-\hbar f_0/\tau_m)-\hbar\omega f_0\,,\label{Eil1a} \\
&&\frac{\Delta_0}{2\pi T}\ln \frac{T_{c0}}{T}=\sum_{\omega >0}
\left (\frac{\Delta_0}{\hbar\omega}-  f_0 \right )\,.
\label{eil4a}
\end{eqnarray}
The first equation here is solved  for $f_0(\Delta_0,\omega)$; \cite{remark} the second gives $\Delta_0(T)$.\cite{Skalski,Ambegaokar,Maki-Parks} 

Eq.\,(\ref{Eil1a}) is used in literature in a different form. Introducing $u=g_0/f_0$ so that $f_0=1/\sqrt{1+u^2}$ and $g_0=u/\sqrt{1+u^2}$ one obtains:\cite{AG,Skalski,Maki-Parks}
\begin{eqnarray}
   \frac{ \hbar\omega }{ \Delta_0} =u    \left(1- \frac{\hbar}{\tau_m\Delta_0\sqrt{1+u^2}}  \right) .
\label{AG-form} 
 \end{eqnarray}
 This equation as well as Eq.\,(\ref{Eil1a})  can be transformed to a quartic equation for $u$ or $f_0 $. \cite{sriva,clem}  The result,
however, is cumbersome and we  prefer to resort to numerical solutions.

Weak supercurrents and fields leave the order parameter modulus unchanged,
but  cause the condensate, i.e., $\Delta$ and  the amplitudes $f$ to acquire an
overall phase $\theta({\bm  r})$. We therefore look for the  perturbed solutions
in the form:
\begin{eqnarray}
\Delta  = \Delta _0 \, e^{i\theta},\qquad 
f =(f_0  +f_1)\,e^{i\theta},\nonumber\\
f^{+} =(f_0  +f_1^+ )e^{-i\theta},\qquad
g =g_0 +g_1       \,      ,
\label{perturbation}
\end{eqnarray}
where  subscripts 1 denote  small corrections.  In the London
limit, the only coordinate dependence is that of the phase $\theta$, i.e.,
        $f_1, f_1^+, g_1 $ can be taken as ${\bm  r}$
independent (taking into account the $\bm r$ dependence of  
$f_1,g_1$ amounts to {\it nonlocal} corrections to the current
response, the question out of the scope of  this paper).\cite{nonloc}  We obtain:  
       \begin{eqnarray}
{\tilde \Delta} g_1 &-&\hbar{\tilde\omega} f_1 =i\hbar f_0{\bm v}{\bm
P}/2\,,\nonumber\\
{\tilde \Delta} g_1 &-&\hbar{\tilde\omega} f^+_1=i\hbar f_0 {\bm
v}{\bm P}/2\,,
\label{system2}\\
2g_0 g_1 &=&-f_0 (f_1 +f^+_1 ) \,,\nonumber
\end{eqnarray}
where
       \begin{eqnarray}
     {\tilde \Delta} = \Delta_0 +\frac{\hbar f_0}{2\tau^-}, \,\,\,\,
     {\tilde\omega} = \omega + \frac{g_0}{2\tau^+} .\qquad 
     \label{w-tilde}
     \end{eqnarray}
 and 
  \begin{eqnarray}
 {\bm P}=\nabla\theta+2\pi{\bm A}/\phi_0  \equiv 2\pi\,
{\bm  a}/\phi_0,
 \label{Pi}
     \end{eqnarray}
is the ``supermomentum" related to   the ``gauge invariant vector potential"  ${\bm  a}$. 
    In writing down the above system, we used the fact that all corrections
are proportional to ${\bm v}{\bm P}$ and their Fermi
surface averages  are zeros. 

To find the current response, we need only $g_1$:
\begin{equation}
g_1=\frac{i\hbar f_0^2 {\bm v}{\bm P}/2}{ {\tilde \Delta} f_0+
     \hbar{\tilde\omega}g_0 }\,.
     \label{g1}
\end{equation} 
The dominator here, 
\begin{eqnarray}
\Delta_0\left(f+g\frac{\hbar\omega}{\Delta_0}+\frac{\hbar}{2\tau\Delta_0}+\frac{\hbar}{2\tau_m\Delta_0}(g^2-f^2)\right),  
\end{eqnarray} 
is manipulated to a simpler form 
 using $\hbar\omega/\Delta_0$ from  Eq.\,(\ref{AG-form})  to obtain:
 \begin{equation}
     {\tilde \Delta} f_0+  \hbar{\tilde\omega}g_0= \Delta_0/f_0 + \hbar/2\tau^-  \,.
     \label{denom}
\end{equation}
Finally, substituting $g_0+g_1$ in the current density (\ref{eil5})  and comparing the result with the London expression  
\begin{equation}
  4\pi {\bm j} /c=-  \lambda^{-2} {\bm a} \,, \label{London}
\end{equation}
we obtain for the penetration depth:
\begin{equation}
 \lambda^{-2}= \frac{16\pi^2 e^2TN(0)v^2}{3c^2}\,  \sum_{\omega}  \frac{f_0^2
}{\Delta_0/f_0+\hbar/2\tau^-}  \,.  \label{lambda-m}
\end{equation}
If only non-magnetic scattering is present, the summand  reduces to
$\Delta_0^2/\beta^2(\beta+\hbar/2\tau)$, $ \beta^2= \Delta_0^2+  \hbar^2\omega^2 $, as it should.\cite{AGD,nonloc,PK-ROPP} 

Thus, the general scheme of  the $\lambda$ evaluation consists of   solving the system of Eqs,\,(\ref{Eil1a}) and (\ref{eil4a}) for $\Delta_0(T)$ and $f_0(\Delta_0,\omega)$ for given scattering parameters;  $\Delta_0(T)$ and $f_0(\Delta_0,\omega)$ then  are substituted in Eq.\,(\ref{lambda-m}) for $\lambda$.

%%%%%%%
\subsection{Numerical procedure}
%%%%%%%

For the numerical work, we introduce dimensionless scattering parameters
       \begin{eqnarray}
 P=\frac{\hbar}{2\pi T_{c0}\,\tau }\,,\qquad P_m=\frac{\hbar}{2\pi T_{c0}\,\tau_m }  \,.  
 \label{P,Pm} 
\end{eqnarray}
The transport scattering parameter $P$ varies between  $0$ and $\infty$. Since   $\tau_m>\tau_{m,crit}=2\hbar/ \Delta_{00}$ where $\Delta_{ 0}(0)$ is the order parameter at $T=0$ in the clean sample, we obtain 
\begin{eqnarray}
 0<P_m<1/4e^\gamma = 0.1404  \,.  
 \label{ Pm} 
\end{eqnarray}
where $\gamma\approx 0.577$ is the Euler constant.  

\begin{figure}[h]
\begin{center}
 \includegraphics[width=8cm] {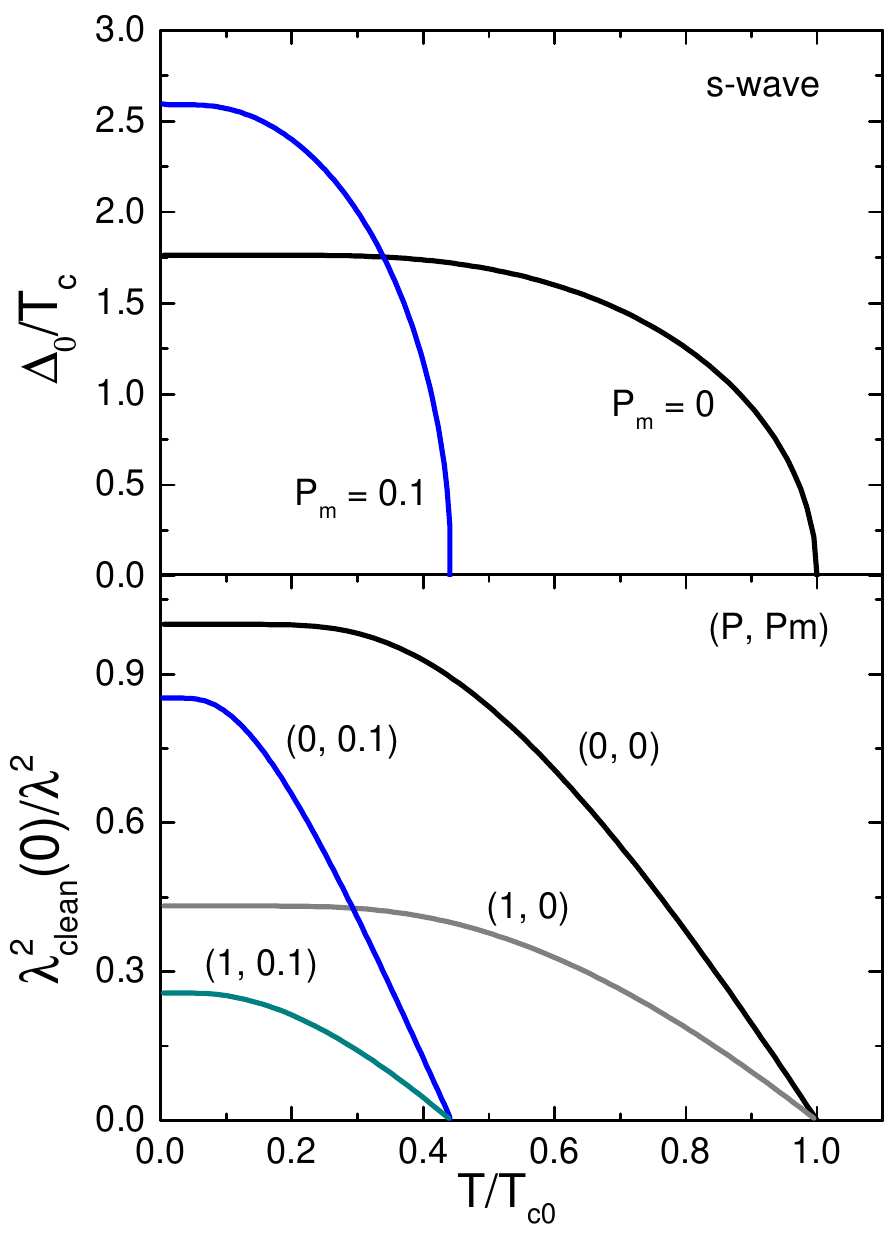}
\caption{The upper panel: numerical solution of Eqs.\,(\ref{Eil1b}) and (\ref{eil4b}) for  $\Delta_0(t)/T_c$ in the absence of magnetic scattering and for the magnetic scattering parameter $P_m=0.1$. 
The lower panel shows     ${\tilde \lambda^{2}}= \lambda ^{2}_{\rm clean}(0)/\lambda^{2}(t)$ for a few combinations of  scattering parameters $P,P_m$.
}
\label{fig1}
\end{center}
\end{figure}

Equations (\ref{Eil1a}) and (\ref{eil4a})  take the form:
      \begin{eqnarray}
&& \sqrt{1-f_0^2}\,(\Delta_1  -  f_0P_m)=t(n+1/2) f_0\,,\label{Eil1b} \\
&& -\ln t=\sum_{n=0}^\infty
\left (\frac{1}{n+1/2}- \frac{t f_0}{\Delta_1 } \right )\,,
\label{eil4b}
\end{eqnarray}
where the reduced temperature and the order parameter are
      \begin{eqnarray}
 t=\frac{T}{T_{c0}}\,,\qquad  \Delta_1 = \frac{\Delta_0(T)}{2\pi T_{c0}}\,.
\label{temp}
\end{eqnarray}
It is worth noting that non-zero solutions of the system (\ref{Eil1b})-(\ref{eil4b}) exist only for $t<t_c=T_c/I_{c0}$ with $t_c$ satisfying the Abrikosov-Gor'kov relation,
      \begin{eqnarray}
   -  \ln t_c =  \psi\left(\frac{1}{2}+\frac{P_m}{t_c}\right)-
\psi\left(\frac{1}{2}\right)\,,
\label{tc}
\end{eqnarray}
where $\psi$ is the digamma function.  In fact, this relation follows from Eq.\,(\ref{Eil1b}) where  $f_0^2$ can be disregarded relative to 1  as $T\to T_c$.  Hence, the numerical solutions of the system (\ref{Eil1b}), (\ref{eil4b})  satisfy $t<t_c$ automatically. 

 Finally, we normalize $\lambda^{-2}$ on the clean limit  $T=0$ value
 \begin{eqnarray}
 \lambda ^{-2}_{\rm clean}(0)= \frac{8\pi   e^2N(0) v^2}{3c^2 } =  \frac{4\pi e^2n}{mc^2 }  
\label{lambda0cl}
\end{eqnarray}
($m$ is the effective mass, $n$ is the carriers density):
\begin{equation}
{\tilde \lambda^{-2}}= \frac{ \lambda^{-2}}{ \lambda ^{-2}_{\rm clean}(0)}=  \sum_{n=0}^\infty  \frac{t\,f_0^3 
}{\Delta_1 +f_0(P-P_m)/2 }  \,.  
\label{tilde-lambda}
\end{equation}

In this work we used Mathematica 9.0 on a HP Z620 Workstation. To obtain the superfluid density, we first solve the system of  Eqs.\,(\ref{Eil1b}) and (\ref{eil4b}) for the Elienberger function $f_0$ and the order parameter $\Delta_1$ thus assuring the self-consistency. Then the penetration depth is found from Eq.\,(\ref{tilde-lambda}). This  simple scheme is very efficient from about $0.1 T_c$ to $T_c$, but may produce artifacts at  lower temperatures  due to a finite upper limit of summations in Eqs.\,(\ref{eil4b}) and (\ref{tilde-lambda}). We therefore use analytic approach for $T=0$, Appendix C, to verify numerical results. The agreement obtained in this manner is shown in the upper panel of Fig.\ref{fig6} where $10^5$ summations and   several hours for a curve of 100 points was needed. Of course, the numerical procedure  can be optimized by employing a temperature dependent upper summation limit.
Representative examples of these calculations are given in Fig\,\ref{fig1}.

%%%%%%%
\subsection{ $\bm {T=0}$}
%%%%%%%

In solving numerically for $\Delta( t)$ and $\lambda(t)$, the low-$T$ region is the most time-consuming. As $t\to 0$, the number of summations needed for reliable numerical results increases. In other words, for  $t=0$ one needs an independent evaluation procedure to confirm general $t$ dependent results. Such a procedure for finding $\Delta(0)$ for a given pair-breaking parameter $P_m$ had, in fact, been given in the  original AG paper.\cite{AG,Maki-Parks} 

To determine $\Delta(0)$ it is convenient to start with the self-consistency equation in the form
      \begin{eqnarray}
 \frac{1}{N(0)V} =2\pi T\sum_{\omega >0} \frac{f_0}{\Delta_0}=\int_0^{\hbar\omega_D} \frac{d\,\hbar\omega}{\Delta_0}f_0 \,.
\label{eil4b}
\end{eqnarray}
where the coupling constant $V$ is related to $T_{c0}$: $\Delta_0(0)=\pi T_{c0} e^{-\gamma}= 2\hbar\omega_D e^{-1/N(0)V}$. 
  
\begin{figure}[ht]
\begin{center}
 \includegraphics[width=8cm] {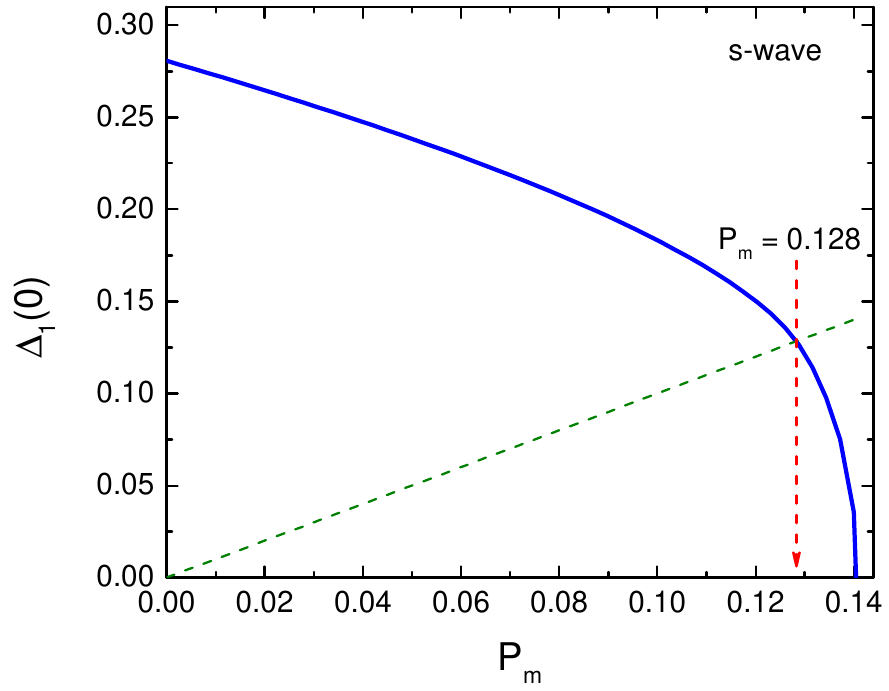}
\caption{(Color online) The zero-$T$ order parameter $\Delta_1=\Delta_0/2\pi T_{c0}$ vs pair-breaking scattering parameter $P_m$ according to  Eq.\,(\ref{system1}).  The vertical dashed line $ P_m=0.128$ separates domains of gapped  and gapless ($0.128<P_m<0.14$) states.  }
\label{fig2}
\end{center}
\end{figure}

We now use Eq.\,(\ref{AG-form}) to replace integration over $\hbar\omega $ in Eq.\,(\ref{eil4b}) with one over $u$. In our notation, the parameter 
 \begin{eqnarray}
\zeta=   \frac{ \hbar }{ \Delta_0\tau_m} =  \frac{P_m}{ \Delta_1}  
\label{zeta} 
 \end{eqnarray}
and the integration     over $u$ goes from $u_1$ to $\hbar\omega_D /\Delta_0$, where $u_1=0$ for $\zeta<1$ and $u_1=\sqrt {\eta_m^2-1}$ for $\zeta>1$.\cite{AG,Maki-Parks}
One then obtains that   
 at $t=0$, the order parameter satisfies the following equations 
 (in our notation):
       \begin{eqnarray}
&&-\ln( 2e^\gamma \Delta_1 )= \frac{\pi}{4}\,\zeta \,, \qquad  \zeta<1\,,\label{system1}\\
&&-\ln( 2e^\gamma \Delta_1 )=  \cosh^{-1}\zeta +\frac{1}{2}\left(\zeta \sin^{-1}\frac{1}{\zeta} -\frac{ \sqrt{\zeta^2-1}}{\zeta}\right)
\,,\nonumber
\end{eqnarray}
where the second line is for $ \zeta>1$. The numerical solution $\Delta_1 (0,P_m)$  of these equations is shown in Fig.\,\ref{fig2}.  
The cross-section of the dashed line $\Delta_1=P_m$ with the curve  $\Delta_1(0,P_m)$ defines the point where  $\zeta=1$. The first of Eqs.\,(\ref{system1}) then gives $P_m =e^{-\gamma-\pi/4}/2\approx 0.128$. This point separates domains of gapped, $0<P_m<0.128$, and gapless, $0.128<P_m<0.14$, states. In fact, the value $0.128/0.14\approx 0.91$ has been established by AG as a fraction of critical density of magnetic impurities where the gap in the electronic spectrum vanishes.

\begin{figure}[htb]
\begin{center}
\includegraphics[width=8cm] {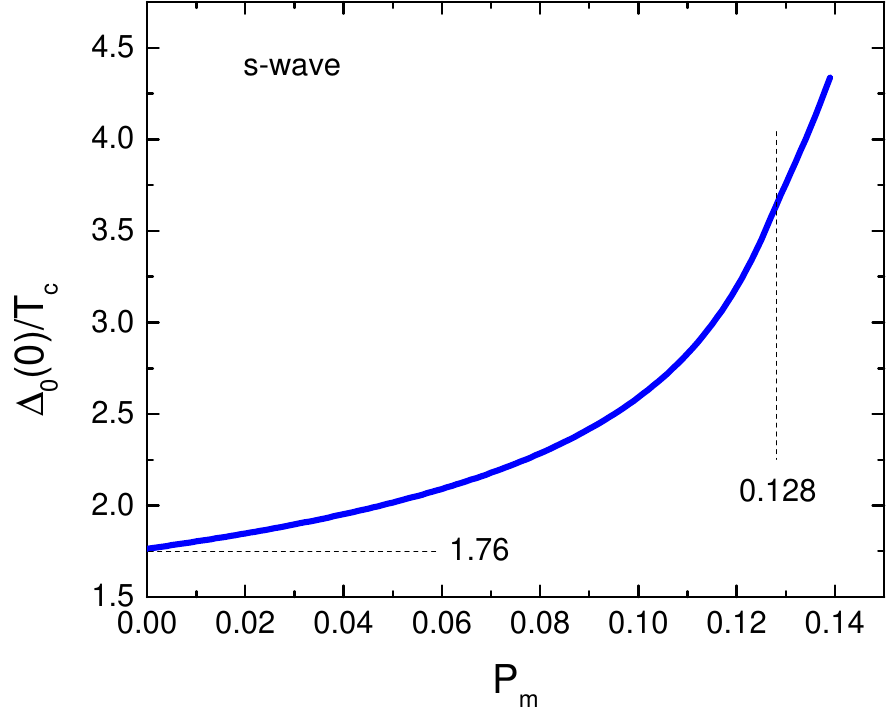}
\caption{  $\Delta_0(0)/T_c$ vs the magnetic scattering parameter $P_m$.    }
\label{fig3}
\end{center}
\end{figure}

\begin{figure}[ht]
\begin{center}
 \includegraphics[width=8cm] {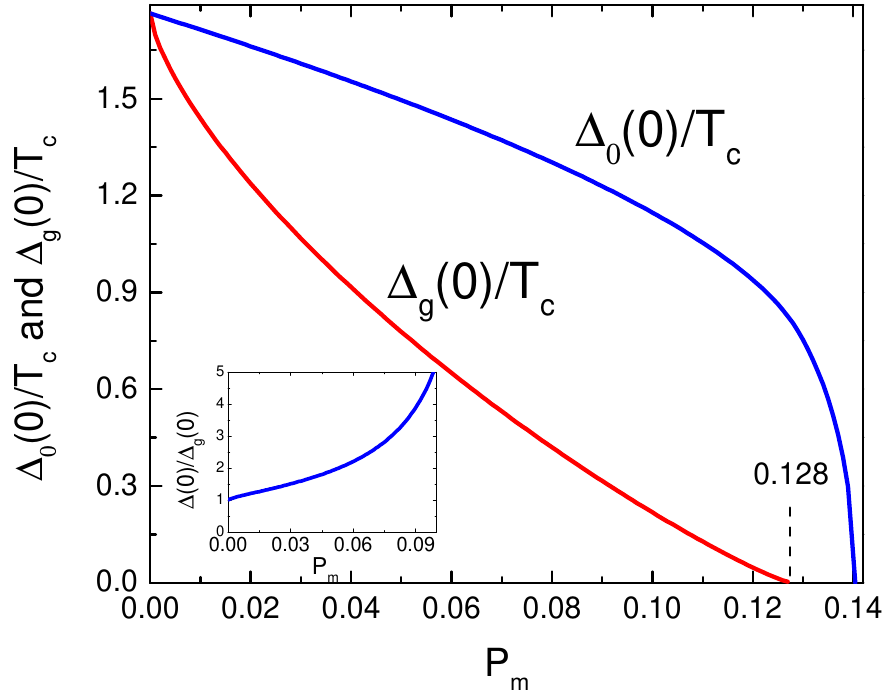}
\caption{(Color online) The ratio of zero-$T$ order parameter  to $T_{c}$ (the upper curve) as compared to the  $T=0$ gap $\Delta_g/T_{c}$ vs pair-breaking scattering parameter $P_m$.  
}
\label{fig4}
\end{center}
\end{figure}

It is instructive to calculate the ratio $\Delta_0(0)/T_c$ as function of   $P_m$,   the quantity often used to identify the superconducting coupling as weak ($\Delta_0(0)/T_c\approx 1.76$) or strong ($\Delta_0(0)/T_c > 1.76$). Fig.\,\ref{fig3} obtained within our weak coupling model shows that the pair breaking interferes with this clear-cut ``weak--strong" distinction.

Given $\Delta_1(0,P_m)$, one can solve  Eq.\,(\ref{Eil1b}) for $f_0$ and  evaluate numerically the penetration depth ${\tilde \lambda^{-2}}(P,P_m)$ at $T=0$ with the help of Eq.\,(\ref{tilde-lambda}). The results are shown in Fig.\,\ref{fig4}. Note that the parameter $P=\pi\xi_0/2e^\gamma\ell\approx 0.88\, \xi_0/ \ell$ so that $P=1$ corresponds to $\xi_0/ \ell \approx 1.1$, i.e., to a quite  clean situation.

Another  point to stress is that calculations of $ \lambda$ involve  the order parameter $\Delta_0$ rather than the gap $\Delta_g$ in the electronic spectrum measured, e.g., in tunneling experiments. In the presence of pair breaking the gap calculated according to AG is $\Delta_g=\Delta_0(1-\zeta^{2/3})^{3/2}$ and it differs from  $\Delta_0 $ for all values of $P_m$ as shown   in Fig.\,\ref{fig4}. 

As mentioned, the calculation of $\lambda $ for $T\to 0$ requires exceedingly large number of summations in Eq.\,(\ref{tilde-lambda}). We verify these results with other method designed for $T=0$ which does not involve summations, Appendix D.

 \begin{figure}[htb]
\begin{center}
  \includegraphics[width=8cm] {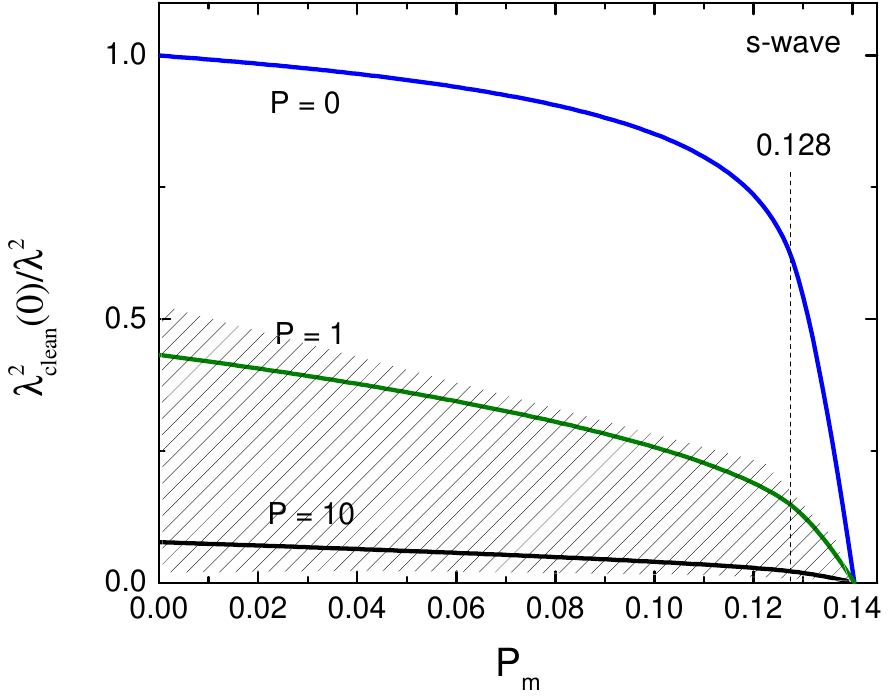}
\caption{(Color online) ${\tilde \lambda^{-2}}(P,P_m)$  at  $T=0$
 vs   $P_m$ for $P=0,  1,  10$. The dashed line at $P_m=0.128$ separates the gapped (left) from gapless (right) domains. Since $P\approx \xi/\ell$, the shaded part roughly corresponds to  scattering parameters for majority of real materials.\cite{Homes,Kogan-Homes}
  }
\label{fig5}
\end{center}
\end{figure}
 
%%%%%%%
\subsection{Strong pair breaking}
%%%%%%%

This is the case when $\tau_m$ is close to $ 2\hbar/\Delta_0(0)$, the
critical value for which $T_c=0$.  According to AG, we have in this domain $\Delta_0^2=2\pi^2(T_c^2-T^2)$ or in our units:
\begin{eqnarray}
       \Delta_1=  (t_c^2-t^2)/2 \,.
\label{Del-a}
\end{eqnarray}
The superconductivity is weak in this domain, $f_0\ll 1$ at all temperatures under $t_c$.\cite{AG} Then, Eq.\,(\ref{Eil1b}) yields in the lowest approximation:
\begin{eqnarray}
       f_0= \frac{\Delta_1}{t(n+1/2)+P_m} \,.
\label{f_GLa}
\end{eqnarray}
Substituting this in Eq,\,(\ref{tilde-lambda}) we obtain:
\begin{eqnarray}
&&{\tilde \lambda^{-2}}= 
 \frac{ 4\Delta_1^2 }{  (P-P_m)^2  }\Big[\psi\left(\frac{P_m}{t}+\frac{1}{2}\right)-\psi\left(\frac{P+P_m}{2t} +\frac{1}{2}\right)\nonumber\\
 &&+\frac{P-P_m}{2 t} \psi^\prime\left(\frac{P_m}{t}+\frac{1}{2}\right)\Big]\,.  
\label{tilde-lam3}
\end{eqnarray}
Note that  $P,P_m$ enter arguments of  $\psi$'s as, e.g.,
\begin{equation}
  \frac {P}{t} =   \frac{\hbar}{2\pi T_c\tau} \gg 1 \,, 
 \label{P/t}
\end{equation}
since $T_c\to 0$ for a strong pair breaking. Hence, we can use large argument asymptotics of functions $\psi$:
\begin{eqnarray}
 {\tilde \lambda^{-2}}=     \frac{2(t_c^2-t^2)} {  (P-P_m)^2  } \left(\frac{P-P_m}{2P_m}-\ln \frac{P+P_m}{2P_m}\right)  ,  
\label{tilde-lam4}
\end{eqnarray}
where the expression (\ref{Del-a}) has been used. 

For the strong pair breaking of interest in this section, $P_m$ is close to the maximum possible value of 0.14 . This implies that $P\gg P_m$ practically for any transport scattering in real materials with     $P\sim\xi_0/\ell $. Expanding Eq.\,(\ref{tilde-lam4}) in small $P_m$, one arrives at:\cite{AG}
\begin{equation}
   {\tilde \lambda^{-2}} = \frac{t_c^2-t^2}{P_mP} \,.
\label{AG_rho}
\end{equation}

One obtains readily for $T=0$ in common units:
\begin{equation}
    \lambda^{-2}(0) = \frac{8\pi^2 }{\hbar c^2P_mT_{c0}} \,\sigma T_c^2\,.
\label{Homes_m}
\end{equation}
where $P_m\approx 0.14$ and $\sigma=2e^2N(0)v^2\tau/3$ is the normal state conductivity.

It is instructive to compare this with  Homes' scaling $\lambda^{-2}(0)   \propto   \sigma T_c$ which works for great many materials.\cite{Homes} This scaling obviously works in the dirty limit where  $\lambda^{-2}(0)   \propto   \sigma \Delta_0(0)$. It has been argued recently\cite{Kogan-Homes} that, in fact, the scaling extends all the way down to $P\approx 1$, i.e., to quite clean situation provided no pair-breaking scattering is present (this also follows from our evaluation of  $\lambda^{-2}(0) $   in Appendix B). 
 Hence we see that when the pair breaking is strong, the Homes scaling is  violated.

%%%%%%
\subsection{Superfluid density}
%%%%%%%

 \begin{figure}[htb]
\begin{center}
 \includegraphics[width=8cm] {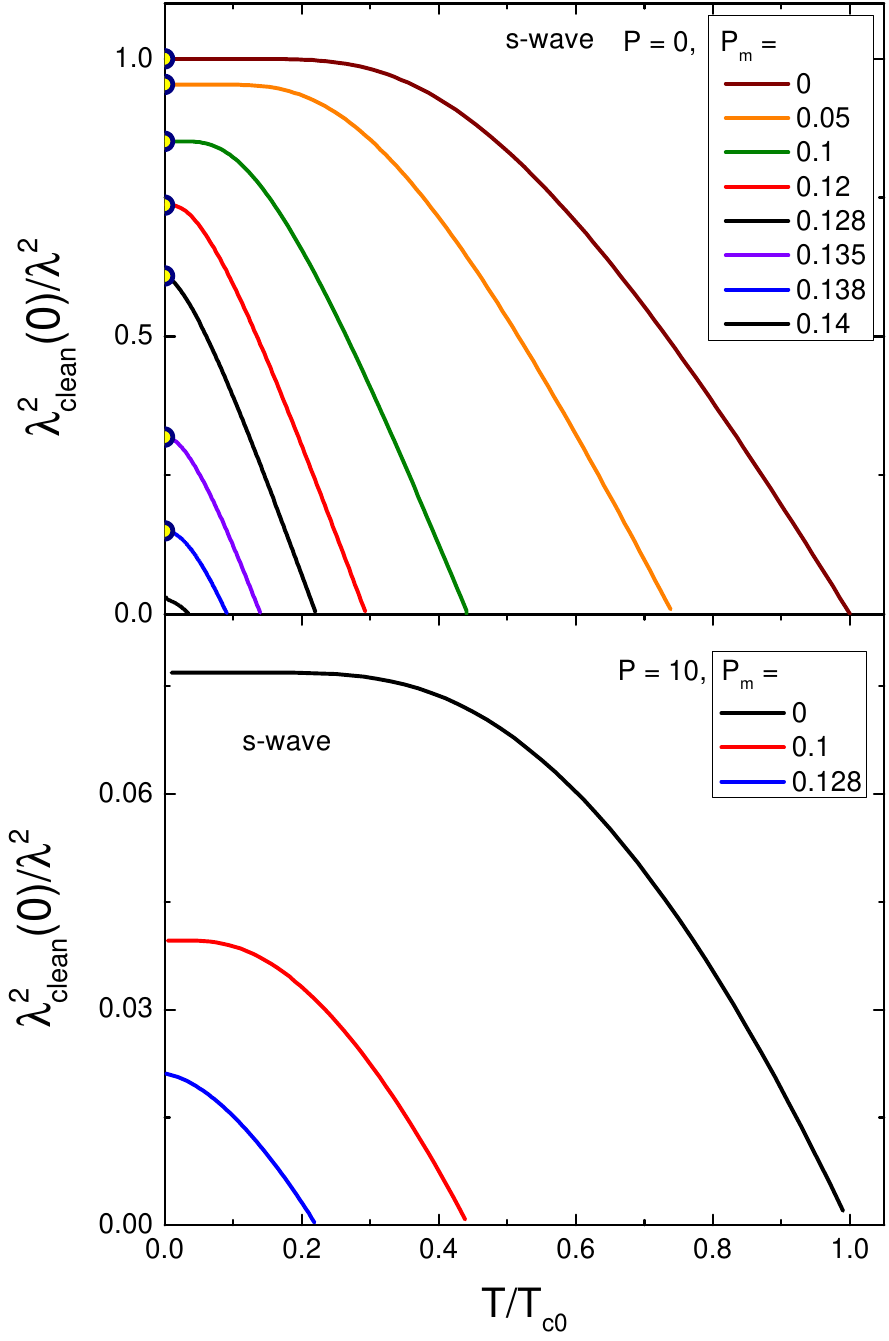}
\caption{(Color online) The superfluid density $\rho $  vs  $T/T_{c0}$ for $P=0$ and a few 
 pair-breaking parameters  $P_m$ shown in the legend.  It is seen that for the gapless state with $P_m>0.128$ the flat low-$T$ part vanishes within the accuracy of this calculation. The dots at $T=0$ are reproduced with the method of Appendix B which does not involve numerical summations.  
 }
\label{fig6}
\end{center}
\end{figure}

   \begin{figure}[h]
\begin{center}
 \includegraphics[width=8cm] {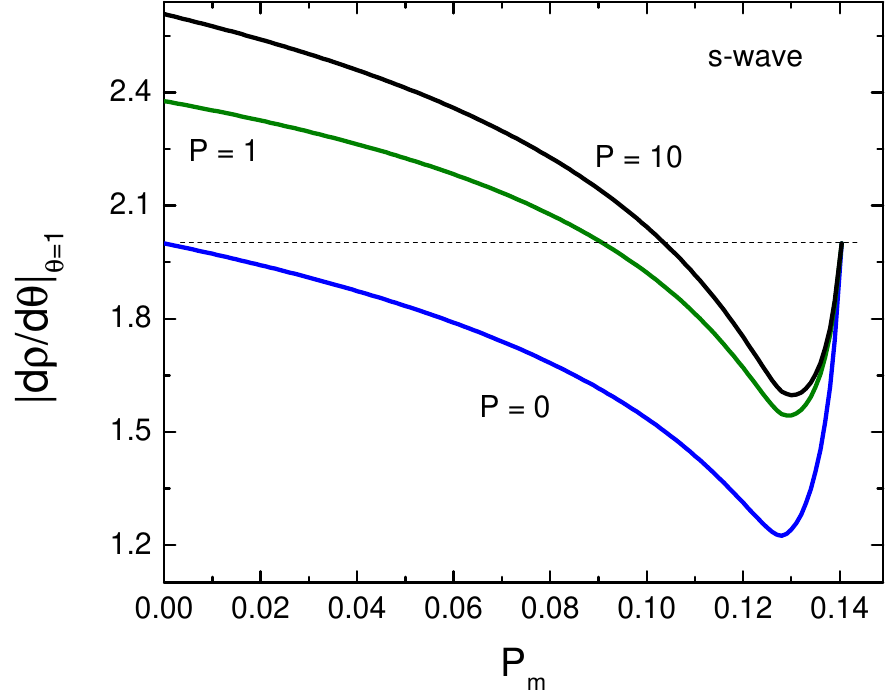}
\caption{(Color online)  The slopes of the normalized superfluid density $|d\rho/d\theta|$ at the phase transition vs the pair-breaking parameter $P_m$ for $P=0, \,\,\,1$, and 10 in the down-up order. Note that in the dirty limit of $P\gg 1$ with no magnetic scattering, $P_m=0$, the slope is 2.66.\cite{Rutgers}  }
\label{fig7}
\end{center}
\end{figure}

It is a common practice to study the normalized superfluid density defined as $\rho(T)=\lambda^2(0)/\lambda^2(T)$ so that $\rho(0)=1$. The pair breaking affects the $T$ dependence of $\rho$ in a dramatic way. Fig.\,\ref{fig6} shows that in the gapless state with $P_m \ge 0.128$ the flat part of $\rho$ as $t\to 0$ nearly disappears and might be confused with the linear d-wave behavior. According to Eq.\,(\ref{tilde-lam4}) it should appear again if $P_m$  approaches the critical value of $0.14$.  

Till now, we have normalized $ \lambda^{-2} (t)$ on the clean limit $ \lambda^{-2}_{\rm clean} (0)$ and employed the reduced temperature $t=T/T_{c0}$. Usually $T_{c0}$ is  unknown and it is preferable to employ the actual $T_c$. Combining the self-consistency Eq.\,(\ref{eil4}) with the AG relation (\ref{tc}) between $T_c$ and $T_{c0}$ one can exclude $T_{c0}$:
      \begin{eqnarray}
   \ln \frac{T_c}{T}&=&\sum_{n=0}^\infty
\left (\frac{1}{n+1/2+\rho_m}- \frac{2\pi T f_0}{\Delta_0 } \right )\,,\label{eil4bb}\\
 \rho_m&=&\frac{\hbar}{2\pi T_c\tau_m}=\frac{P_m}{t_c}\,.
\label{rm}
\end{eqnarray}

 We now focus on the slope  $d \rho/dT$ at $T_c$. To find this quantity  we need to solve the  self-consistency equation as $T\to T_c$ where both  $\Delta_0$ and $f_0$ go to zero. We look for solutions of   Eq.\,(\ref{Eil1a}) in the form $f_0=f_1+f_2$, $f_2\ll f_1\ll 1$, to obtain
      \begin{eqnarray}
  f_0=\frac{\Delta_0}{\hbar\omega_m}- \frac{\omega}{2\omega_m}\, \frac{\Delta_0^3}{\hbar^3\omega_m^3}\,,\qquad \omega_m = \omega+\frac{1}{\tau_m}\,.
  \label{f1f2}
\end{eqnarray}
Substitute this in Eq.\,(\ref{eil4bb}) and do  the summation:
     \begin{eqnarray}
 \Delta_0^2 =  \frac{16\pi^2T_c^2[\rho_m\psi^\prime(\rho_m +1/2 )-1]}{\psi^{\prime\prime}\left(\rho_m +1/2\right)+\frac{\rho_m}{3}\psi^{\prime\prime\prime}\left(\rho_m +1/2\right)} (1-\theta)\,,\qquad
  \label{D(Tc)}
\end{eqnarray}
where $\theta=T/T_c$ (not to confuse with $t=T/T_{c0}$).  

In evaluation of  $\lambda^{-2}$ of Eq.\,(\ref{lambda-m}) near $T_c$, the first term in the expansion  (\ref{f1f2}) suffices. After simple algebra we obtain:
     \begin{eqnarray}
&&\tilde \lambda^{-2}= \frac{\Delta_0^2}{4\pi^2 T_c^2}  \sum_n  \left(n+\frac{1}{2}+\rho_m\right)^{-2} \left(n+\frac{1}{2}+\frac{\rho^+}{2}\right)^{-1} \nonumber\\
&&=\frac{\Delta_0^2}{2\pi^2 T_c^2\rho^2_-}\Big[2\psi \left(\rho_m +\frac{1}{2}\right)-2\psi \left(\frac{\rho^+ +1}{2}\right)\nonumber\\
&&+\rho^-\psi^\prime \left(\rho_m +\frac{1}{2}\right)\Big],\quad \rho^\pm=\frac{\hbar}{2\pi T_c\tau^\pm}=\frac{P\pm P_m}{t_c } \,.
 \label{lambda-Tc}
\end{eqnarray}
Combining this with Eq.\,(\ref{D(Tc)}) for $\Delta_0$ near $T_c$ and utilizing    $\tilde\lambda^{-2}(0)$ calculated above we obtain the slope of the normalized superfluid density $ d\rho/d\theta=(d\lambda^{-2}/d\theta)/\lambda^{-2}(0)$ at the transition, $\theta=1$. 

Results of this evaluation are shown in Fig.\,\ref{fig7}. Main features of these curves  are: (i) with no magnetic scattering, $P_m=0$, the slope $|d\rho/d\theta|$ increases with increasing transport scattering $P$ from the clean limit value of 2 up to the dirty limit 2.66,\cite{Rutgers} (ii) with increasing $P_m$ the slopes decrease and reach minimum near the boundary between gapped and gapless states at  $P_m= 0.128$,  (iii) in the gapless domain  $0.128 < P_m < 0.14 $, the slopes increase and tend to the value of 2, in agreement with AG prediction for this limit.\cite{AG}

%%%%%%%
\section{d -wave}
%%%%%%%

   \begin{figure}[h]
\begin{center}
\includegraphics[width=8cm] {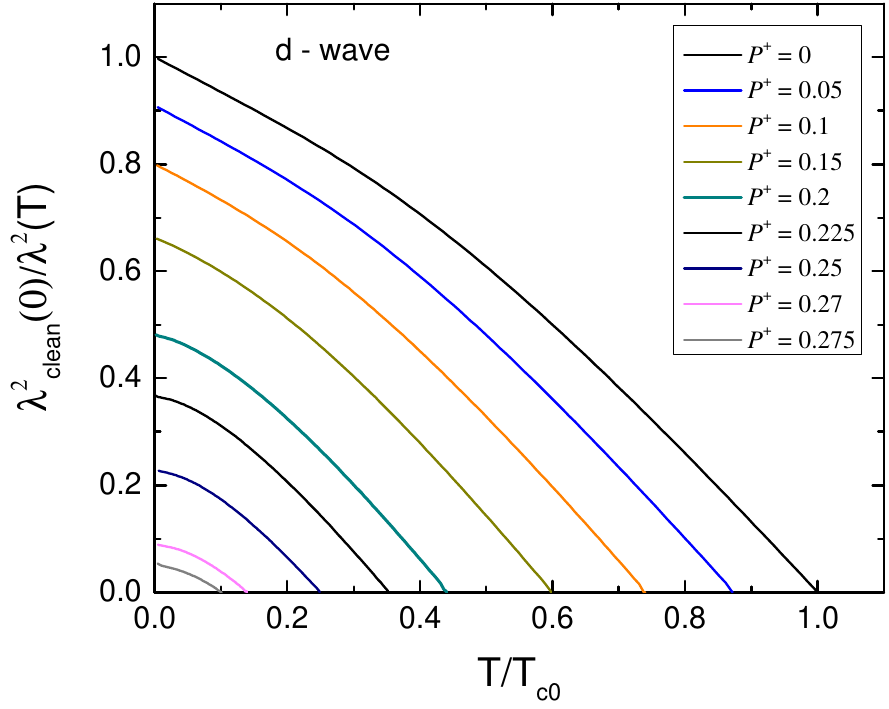}
\caption{(Color online)  The superfluid density of d-wave superconductors   for    a few values of the combined scattering parameter $P^+= P+P_m=\hbar/2\pi T_{c0}\tau^+$ shown in the legend; $P^+_{crit}=0.128$ corresponds to $T_c=0$.  }
\label{fig8}
\end{center}
\end{figure}

Equations (\ref{Eil1})-(\ref{eil3}) hold for any anisotropic order parameter. We assume a factorizable form of the coupling potential responsible for superconductivity $V({\bm k},{\bm k^\prime})= V_0\Omega({\bm k})\Omega({\bm k^\prime})$ and of the order parameter 
 $\Delta({\bm k},{\bm r},T)=  \Omega({\bm k})\Psi({\bm r },T)$. The self-consistency equation for the uniform state then takes the form:
\begin{eqnarray}
   \frac{\Psi_0}{2\pi T}\ln \frac{T_{c0}}{T}=\sum_{\omega >0}
\left (\frac{\Psi_0}{\hbar\omega}-\Big\langle \Omega f\Big\rangle\right )\,.
\label{eil4d} 
 \end{eqnarray}
The function   $  \Omega({\bm k}) $ determines the dependence of $\Delta $ on the position at the Fermi surface and is normalized: $\langle \Omega  \rangle^2=1$. 
For the  Fermi surface as a rotational ellipsoid, with nodes of the d-wave order parameter  along meridians, $ \Omega   =\sqrt{2}\cos 2\varphi$ where $\varphi$ is the azymuth.\cite{KP-ROPP}   
 
For the  field-free state we average Eq.\,(\ref{Eil1}) over the Fermi surface to obtain $\langle f_0  \rangle=0$ so that we have:
\begin{eqnarray}
\Delta_0 g_0-\hbar{\tilde\omega}  f_0=0\,,\quad {\tilde\omega} = \omega+\frac{G}{2\tau^+}\,,\quad G=\langle g_0\rangle.\qquad
\label{Eil0d} 
 \end{eqnarray}
Together with Eq.\,(\ref{eil3}) this gives;
\begin{eqnarray}
f_0=\Delta_0 /{\tilde\beta}\,,\quad  g_0=\hbar{\tilde\omega}/{\tilde\beta}\,,\quad {\tilde\beta}^2= \Delta_0^2+\hbar^2{\tilde\omega}^2.
\label{f0d} 
 \end{eqnarray}
 
 Since $\tilde\omega$ does not depend on the angle $\varphi$, we have 
\begin{eqnarray}
 G&=&\hbar \tilde\omega\left\langle\frac{1} {  \tilde\beta}\right \rangle =\frac{\hbar\tilde\omega}{2\pi}\int_0^{2\pi}\frac{d\varphi}{\sqrt{2\Psi_0^2\cos^22\varphi+\hbar^2\tilde\omega^2}}\nonumber\\
&=&\frac{2 \hbar\tilde\omega}{\pi \sqrt{2\Psi_0^2 +\hbar^2\tilde\omega^2}} {\bm K}\left(\frac{2\Psi_0^2}{2\Psi_0^2 +\hbar^2\tilde\omega^2}\right),
\label{<g>} 
 \end{eqnarray}
where ${\bm K}(m)$ is the Complete Elliptic Integral with $m= 2\Psi_0^2/(2\Psi_0^2 +\hbar^2\tilde\omega^2)$.\cite{Abr}  This equation can be solved numerically to find $G$ for given $t,n,\Psi_0,$ and the scattering rate $1/\tau^+$. Taking $2\pi T_{c0}$ as a unit of energy, we obtain this equation in dimensionless form:
\begin{eqnarray}
 G&=& \frac{2\tilde\omega_1}{\pi \sqrt{2\Psi_1^2 + \tilde\omega_1^2}} {\bm K}\left(\frac{2\Psi_1^2}{2\Psi_1^2 + \tilde\omega_1^2}\right),\label{G}\\
\Psi_1 &=&\frac{\Psi_0}{2\pi T_{c0}},\qquad P^+= \frac{\hbar}{2\pi T_{c0}\tau^+}\,,\label{Psi-p}\\
  \tilde\omega_1&=&\frac{\hbar\tilde\omega}{2\pi T_{c0}}=t\left(n+\frac{1}{2}\right)+\frac{GP^+}{2}\,. 
\label{variables} 
 \end{eqnarray}

After averaging over the Fermi surface, the self-consistency equation (\ref{eil4d}) takes the dimensionless form:
\begin{eqnarray}
&&  \sum_{n=0}^\infty \left[\frac{1}{n+1/2} 
-\frac{2 t \sqrt{2}  }{\pi\Psi_1} \,\frac{\bm E(m )-(1-m )\bm K(m )}{ \sqrt{m }} \right]\qquad\nonumber\\
&&= -\ln t\,,\qquad\qquad  
m = \frac{ 2\Psi_1^2}{2\Psi_1^2+ \tilde\omega_1^2}  \,. \qquad \label{eil4dd}
 \end{eqnarray}
The system of Eqs.\,(\ref{G})--(\ref{eil4dd}) is solved numerically to obtain the order parameter $\Psi_1=\Delta_{1,max}/\sqrt{2}$.

It is worth noting that for the d-wave symmetry, the transport and magnetic Born scattering enter the self-consistency Eq.\,(\ref{eil4dd}) only additively. Hence, both rates affect the order parameter and, in particular, the critical temperature depression in exactly the same manner. Formally, this means that instead of two scattering parameters, $P$ and $P_m$, one has only one $P^+=P+P_m$, which simplifies  treatment of the d-wave case as compared to the s-wave. It remains to be seen whether or not this feature still holds for other than Born scattering regimes.

The  perturbation procedure, as described for the s-wave, yields  the correction $g_1$ to $g_0$ in the presence of weak fields: 
\begin{equation}
g_1=\frac{i\hbar f_0^2 {\bm v}{\bm P}/2}{  \Delta_0 f_0+
     \hbar{\tilde\omega}g_0 } = \frac{i\hbar \Delta_0^2 {\bm v}{\bm P} }{ 2 \tilde\beta^3   }\,.
     \label{g1d}
\end{equation} 
As was done above, one substitutes this in the expression (\ref{eil4}) for the current density and compares the result  with the anisotropic version of London Eq.\,(\ref{London}) to get the penetration depth:
\begin{equation}
( \lambda^2)_{ik}^{-1}= \frac{16\pi^2 e^2TN(0) }{ c^2}\,  \sum_{\omega} \left\langle \frac{v_iv_k \Delta_0^2}{\tilde\beta^3 } \right\rangle \,.  
\label{lambda-d}
\end{equation}

At first sight, for the d-wave order parameter, $\lambda$ can be anisotropic even on the Fermi sphere. This, however,  is not the case: 
\begin{eqnarray}
   ( \lambda^2)_{aa}^{-1}&\propto&   \left\langle \frac{v_a^2 \Delta_0^2}{\tilde\beta^3 } \right\rangle 
   =  \left\langle \sin^2\theta\cos^2\varphi \,\Phi(\cos^22\varphi) \right \rangle, \qquad\nonumber\\  
    ( \lambda^2)_{cc}^{-1}&\propto &   \left\langle \cos^2\theta\,\Phi(\cos^22\varphi) \right \rangle, \qquad \Phi=\Delta_0^2/\tilde\beta^3\,,
     \end{eqnarray}
  which are easily shown to be the same. 
Hence, the tensor $  (\lambda^2)^{-1} _{ik}$ is reduced to $\lambda ^{-2}\delta _{ik}$. 
Apparently, this is the property of the order parameter $\Delta\propto (k_x^2- k_y^2)$ on the Fermi sphere.  In particular, this means that for a d-wave order parameter on a Fermi sphere, $\lambda_{ab}(T)=\lambda_c(T)$  for any Born scattering, either transport or magnetic. 
However amusing this conclusion is, it suggests that the contribution of the d-wave {\it per se} to the $\lambda$ anisotropy is weak relative to the contribution of anisotropic Fermi surfaces. 
It should be  noted here  that Ref.\,\onlinecite{Muz89} concludes that the order parameter of the form $(k_x+i k_y)^2$, a mixture of two d-waves, does produce  anisotropy of $\lambda$ even if the Fermi surface is a sphere. 
   
We normalize $ \lambda^{-2} $ on $ \lambda ^{-2}_{\rm clean}(0)$ of Eq.\,(\ref{lambda0cl}) and obtain after performing the Fermi sphere average:
\begin{eqnarray}
  \tilde\lambda  ^{-2}&=& \frac{ t\sqrt{ 2} }{ \pi\Psi_1}  \sum_{n=0}^\infty \sqrt{m }  \left[ \bm K(m ) -   \bm E(m )  \right] .
\label{lambda-cc} 
\end{eqnarray}
  This relation for $T=0$ has been given in Ref.\,\onlinecite{Muz88}.

We note that  for $P^+=0$, $\lambda(0)$ coincides with  $ \lambda _{\rm clean}(0)$   in agreement with the general argument based on Galilean invariance: in the absence of scattering at $T=0$ all carriers take part in the supercurrent independently of the order parameter value or its symmetry. 
 
   \begin{figure}[h]
\begin{center}
\includegraphics[width=8cm] {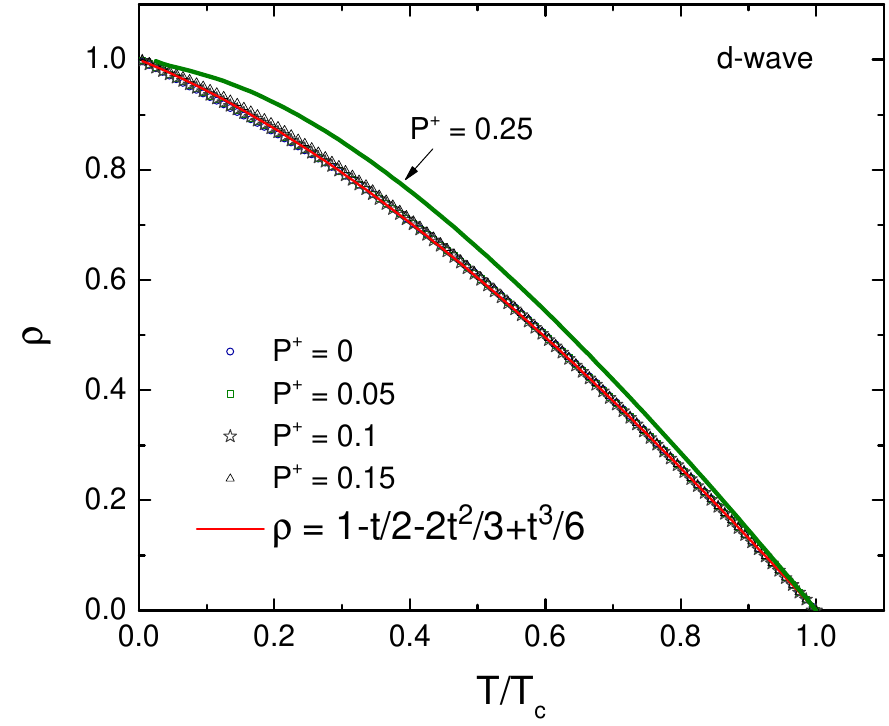}
\caption{(Color online)  The superfluid density of d-wave superconductors vs reduced temperature $T/T_c$  for  a few values of $P^+$ up to $0.15$ which corresponds to $T_c/T_{c0}\approx 0.6$. A simple polynomial fit gives a good approximation of the curves presented.  
}
\label{fig9}
\end{center}
\end{figure}

   \begin{figure}[h]
\begin{center}
\includegraphics[width=8cm] {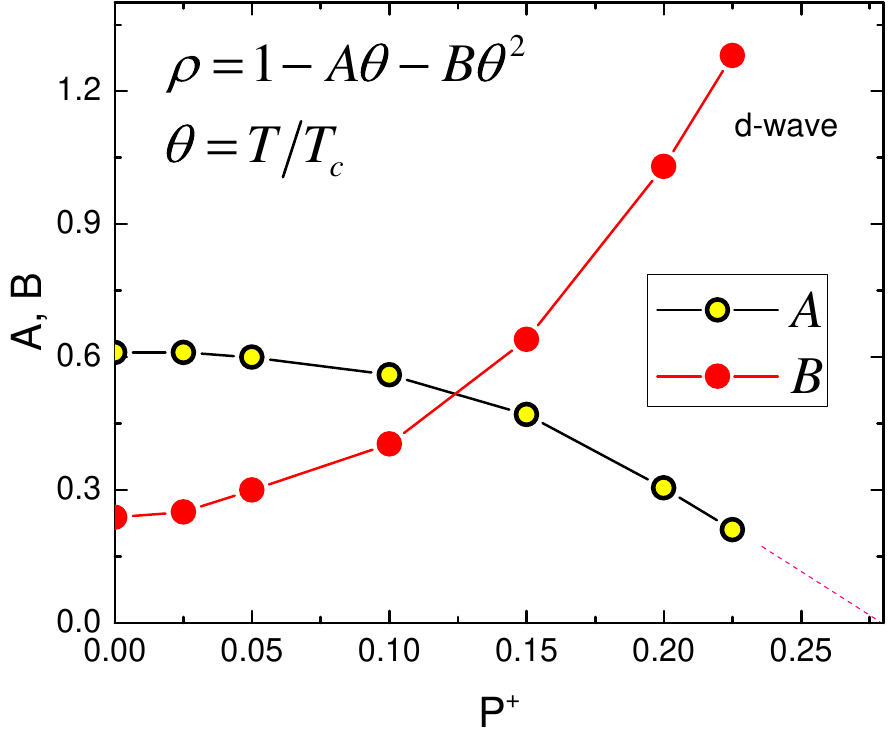}
\caption{(Color online)  Fit of the superfluid density $\rho(T/T_c)$ of d-wave superconductors   to a square polynomial  in the interval $0<\theta < 0.3$ showing relative contributions of linear and quadratic terms with increasing pair-breaking parameter $P^+$. Although as $\theta\to 0$,  $B\theta^2/A\theta\to 0$, the coefficient $B$   
remains comparable to $A$.
}
\label{fig10}
\end{center}
\end{figure}

Figure 9 shows the normalized superfluid density $\rho=\lambda^{-2}(T)/\lambda^{-2}(0)$  calculated numerically versus reduced temperature $T/T_c$ for a few scattering parameters $P^+$. A remarkable feature to note: all curves with $P^+$ up to about half of the maximum possible value of 0.28 are nearly the same. In particular they have the slope  at $T_c$ close to the   clean limit value of 4/3. Example of deviations from this nearly universal form for $P^+=0.25$ is also shown. We conclude again that the clean limit d-wave form of $\rho(T/T_c)$ is only weakly sensitive   to the Born scattering.

\section{Discussion}
 
 We have studied effects of the  transport and pair-breaking scattering  in the Born approximation upon temperature dependence of the penetration depth for s- and d-wave order parameters on isotropic Fermi surfaces.  In practice of analyzing $\lambda(T)$ data,  our work may prove useful since it shows that the pair-breaking scattering changes even a qualitative character of $\lambda(T)$ curves. Examples of Fig.\,\ref{fig6}
for the s-wave case demonstrate  clearly that a sufficiently strong pair breaking practically eliminates the flat low-temperature part of superfluid density curves and makes them qualitatively similar to the d-wave linear behavior. 

For the d-wave symmetry we find nearly universal behavior of the normalized superfluid density $\rho(\theta)=\lambda^2(0)/\lambda^2(\theta)$ ($\theta=T/T_c$,   not to confuse with $t=T/T_{c0}$) for   $P^+=P+P_m$ up to $\approx 0.15$ (whereas $P^+_{crit}=0.28$ kills superconductivity altogether). 

\begin{figure}[htb]
\begin{center}
\includegraphics[width=8cm] {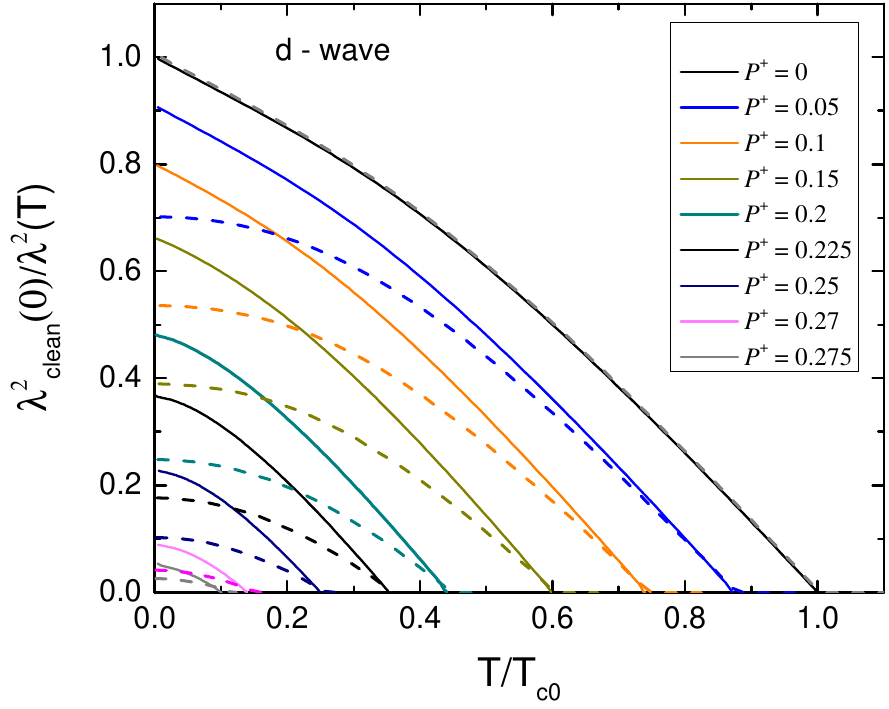}
\caption{(Color online)  Dashed curves: the superfluid density  for   scattering parameters $P^+ $ of Fig.\,\ref{fig8} in the unitary limit for d-wave superconductors.    }
\label{fig11}
\end{center}
\end{figure}

A note of caution: we consider the scattering in the Born approximation which, of course, restricts applicability of our results.  To demonstrate how strong the effect of the scattering approximation might be we show in Fig.\,\ref{fig11} the results for the superfluid density of a d-wave superconductor calculated for a unitary scattering limit for the same input parameters as those of Fig.\,\ref{fig8}. One can see that even a weak scattering eliminates the linear low temperature signature of the d-wave order parameter and transforms it in the $T^2$ behavior in agreement with early results.\cite{Hirsch-Gold,Carbotte}   

 However, when confronted with data interpretation  on new materials, one never knows up front what kind of scattering model should be employed, so that it is reasonable to start with the simplest situation of the Born approximation.   Discussion of the pair-breaking scattering effects within the t-matrix approach and in the unitary limit was a subject of a number of excellent theoretical papers;\cite{Muz88,Muz89,Muz94,Hirsch-Gold,Sun-Maki,Sauls,Carbotte} still, a number of issues there related to the data interpretation  deserve further study and will be considered elsewhere. 

\section{ACKNOWLEDGMENTS}
We are grateful to P. Hirschfeld,  J. Clem, and M. Tanatar for illuminating discussions.   
The work at Ames Lab was supported by the U.S. Department of Energy, Office of Basic Energy Sciences, Division of Materials Sciences and Engineering under contract No. DE-AC02-07CH11358. VM acknowledges support from the Center for Emergent
Superconductivity, an Energy Frontier Research Center funded by the US DOE, Office of Science, under Award No. DE-AC0298CH1088.

\appendix

\section{Notation}

We use a number of dimensional  and reduced quantities. For readers convenience  we provide a short list below:

  $T_{c0}$ is the critical temperature in absence of pair-breaking  scattering. 

   The reduced temperatures are $t=T/T_{c0}$ and $\theta=T/T_{c}$.

 $\Delta(T,\bm r)$  is the order parameter with the dimension of  energy. $\Delta_0(T)$ with the dimension of  energy  is the order parameter of the uniform field-free state.
$\Delta_1(t)= \Delta_0(T)/2\pi T_{c0}$. 

For the d-wave,  
\begin{eqnarray} 
&&\Delta_0(T)=\Delta_{0,max}\cos 2\varphi=\Psi_0 \sqrt{2} \cos 2\varphi\,,\nonumber\\
&& \Delta_{1,max}=\frac{\Delta_{0,max}}{2\pi T_{c0}}=\frac{\Psi_{0,max}\sqrt{2}}{2\pi T_{c0}}=\Psi_1 \sqrt{2}\,. \nonumber
\end{eqnarray}

%%%%%%%
\section{$\lambda(0)$}
%%%%%%%

We offer here a way of evaluating $\lambda(0)$ which does not involve summations and can be used to verify $\lambda(T)$ obtained with the help of Eq.\,(\ref{tilde-lambda}).
At $T=0$ Eq.\,(\ref{lambda-m}) gives for the isotropic s-wave:
\begin{equation}
 \tilde{}\lambda^{-2} =     \int_0^{\hbar\omega_D}  \frac{d(\hbar\omega)/\Delta_0}{(1+u^2)(\sqrt{1+u^2}+\eta_-)}  \,,
 \label{eq22}
\end{equation}
where $1+u^2=1/f_0^2$ and 
\begin{equation}
  \eta_-=\frac{\hbar}{2\tau^-\Delta_0}=\frac{P-P_m}{2 \Delta_1}  \, 
 \label{eq23}
\end{equation}
is the relevant scattering parameter. Integration here can be done by going to the variable $u$ as explained in derivation of Eq.\,(\ref{system1}). 
For $\zeta =P_m/\Delta_1<1$ the integration over $u$ is from $0$ to $\hbar\omega_D/\Delta_0$ and we obtain: 
       \begin{eqnarray}
&&{\cal I} = \frac{\pi}{2\eta_-}\left(1-\frac{4 \tan^{-1}\frac{1-\eta_- }{\sqrt{1-\eta_-^2}}}{\pi \sqrt{1-\eta_-^2}}\right)\nonumber\\
&&+\zeta \frac {12\eta_- +8\eta_-^3-3\pi(2+\eta_-^2)} {12 \eta_-^4} \label{eq26} \\ 
&&-\frac{2\zeta }{\eta_-^4\sqrt{1-\eta_-^2}}
  \tan^{-1}\frac{\eta_- -1 }{\sqrt{1-\eta_-^2}} \,. \nonumber
\end{eqnarray}
For purely transport scattering, $\zeta=0$, this reduces to the result of Ref.\,\onlinecite{Kogan-Homes}. This expression, in fact, covers arbitrary transport scattering and the pair breaking up to $P_m=0.128$ corresponding to a strong suppression of the critical temperature $T_c/T_{c0}=0.22$. 

In the gapless state with $\zeta >1$, the integral over $u$ is from $\sqrt{\zeta^2-1}$ to $\hbar\omega_D/\Delta_0$. The integration is doable analytically,  
but the result is very cumbersome and not really illuminating. One can easily do the integration numerically.

\end{document}